# Irregular Bipolar Fuzzy Graphs

**Sovan Samanta**
ssamantavu@gmail.com

**Madhumangal Pal**
mmpalvu@gmail.com

Department of Applied Mathematics with Oceanology and Computer Programming,
Vidyasagar University, Midnapore - 721 102, India.

### Abstract

*In this paper, we define irregular bipolar fuzzy graphs and its various classifications. Size of regular bipolar fuzzy graphs is derived. The relation between highly and neighbourly irregular bipolar fuzzy graphs are established. Some basic theorems related to the stated graphs have also been presented.*

## 1 Introduction

In 1965, Zadeh [25] introduced the notion of fuzzy subset of a set as a method of presenting uncertainty. The fuzzy systems have been used with success in last years, in problems that involve the approximate reasoning. It has become a vast research area in different disciplines including medical and life sciences, management sciences, social sciences, engineering, statistics, graph theory, artificial intelligence, signal processing, multiagent systems, pattern recognition, robotics, computer networks, expert systems, decision making, automata theory etc.

In 1975, Rosenfeld [20] introduced the concept of fuzzy graphs. In 1994, Zhang [26] initiated the concept of bipolar fuzzy sets as a generalization of fuzzy sets. Bipolar fuzzy sets are extension of fuzzy sets whose range of membership degree is $[-1,1]$. In bipolar fuzzy set, membership degree $0$ of an element means that the element is irrelevant to the corresponding property, the membership degree within $(0,1]$ of an element indicates that the element somewhat satisfies the property, and the membership degree within $[-1,0)$ of an element indicates the element somewhat satisfies the implicit counter property. For example, sweetness of foods is a bipolar fuzzy set. If sweetness of foods has been given as positive membership values then bitterness foods is for negative membership values. Other tastes like salty, sour, pungent (e.g. chili), etc are irrelevant to the corresponding property. So these foods are taken as zero membership values.

In 2011, Akram [2] introduced the concept of bipolar fuzzy graphs and defined different operations on it. Bipolar fuzzy graph theory is now growing and expanding its applications. The theoretical developments in this area is discussed here.

### 1.1 Review of literature

After Rosenfeld [20], fuzzy graph theory is increased with a large number of branches. Samanta and Pal introduced fuzzy tolerance graphs [21], fuzzy threshold graphs [22], fuzzy competition graphs [23] and bipolar fuzzy hypergraphs [24]. Mathew and Sunitha [6] described the types of arcs in a fuzzy graph. Nagoorgani and Malarvizhi [12] established the isomorphism properties of strong fuzzy graphs. Nagoorgani and Radha [11] defined regular fuzzy graphs. Nagoorgani and Vadivel [13] showed relations between the parameters of independent domination and irredundance in fuzzy graphs. Nagoorgani and Vijayalaakshmi [14] defined insentive arc in domination of fuzzy graph. Nair and Cheng [17] defined cliques and fuzzy cliques in fuzzy graphs. Nair [18] established the definition of perfect and precisely perfect fuzzy graphs. Natarajan [19] et al. showed strong (weak) domination in fuzzy graphs. Akram [1] defined different operations on bipolar fuzzy graphs. He also shown that the automorphism property of bipolar fuzzy graph. Strong bipolar fuzzy graphs were also introduced here. He also introduced regular bipolar fuzzy graphs [2].

## 2 Preliminaries

A *fuzzy set* $A$ on a set $X$ is characterized by a mapping $m: X \to [0,1]$, called the membership function. A fuzzy set is denoted as $A = (X, m)$. A *fuzzy graph* [20] $\xi = (V, \sigma, \mu)$ is a non-empty set $V$ together with a pair of functions $\sigma : V \to [0,1]$ and $\mu : V \times V \to [0,1]$ such that for all $u, v \in V$, $\mu(u,v) \leq \sigma(u) \wedge \sigma(v)$ (here $x \wedge y$ denotes the minimum of $x$ and $y$). *Partial fuzzy subgraph* $\xi' = (V, \tau, \nu)$ of $\xi$ is such that $\tau(v) \leq \sigma(v)$ for all $v \in V$ and $\mu(u,v) \leq \nu(u,v)$ for all $u, v \in V$. *Fuzzy subgraph* [8] $\xi'' = (P, \sigma', \mu')$ of $\xi$ is such that $P \subseteq V$, $\sigma'(u) = \sigma(u)$ for all $u \in P$, $\mu'(u,v) = \mu(u,v)$ for all $u, v \in P$. A fuzzy graph is *complete* [11] if $\mu(u,v) = \sigma(u) \wedge \sigma(v)$ for all $u, v \in V$. The *degree of vertex* $u$ is $d(u) = \sum_{(u,v) \in \xi} \mu(u,v)$. The *minimum degree* of $\xi$ is $\delta(\xi) = \wedge \{d(u) | u \in V\}$. The *maximum degree* of $\xi$ is $\Delta(\xi) = \vee \{d(u) | u \in V\}$. The *total degree* [11] of a vertex $u \in V$ is $td(u) = d(u) + \sigma(u)$. A fuzzy graph $\xi = (V, \sigma, \mu)$ is said to be *regular* [11] if $d(v) = k$, a positive real number, for all $v \in V$. If each vertex of $\xi$ has same total degree $k$, then $\xi$ is said to be a *totally regular* fuzzy graph. A fuzzy graph is said to be *irregular* [16], if there is a vertex which is adjacent to vertices with distinct degrees. A fuzzy graph is said to be *neighbourly irregular* [16], if every two adjacent vertices of the graph have different degrees. A fuzzy graph is said to be *totally irregular*, if there is a vertex which is adjacent to vertices with distinct total degrees. If every two adjacent vertices have distinct total degrees of a fuzzy graph then it is called *neighbourly total irregular* [16]. A fuzzy graph is called *highly irregular* [16] if every vertex of $G$ is adjacent to vertices with distinct degrees. The *complement* [8] of fuzzy graph $\xi = (V, \sigma, \mu)$ is the fuzzy graph $\xi' = (V, \sigma', \mu')$ where $\sigma'(u) = \sigma(u)$ for all $u \in V$ and

$$\mu'(u,v) = \begin{cases} 0, & \text{if } \mu(u,v) > 0, \\ \sigma(u) \wedge \sigma(v), & \text{otherwise.} \end{cases}$$

Let $X$ be a nonempty set. A *bipolar fuzzy set* [26] $B$ on $X$ is an object having the form $B = \{(x, m^+(x), m^-(x)) \mid x \in X\}$, where $m^+ : X \to [0,1]$ and $m^- : X \to [-1,0]$ are mappings. If $m^+(x) \neq 0$ and $m^-(x) = 0$, it is the situation that $x$ is regarded as having only positive satisfaction for $B$. If $m^+(x) = 0$ and $m^-(x) \neq 0$, it is the situation that $x$ does not satisfy the property of $B$ but somewhat satisfies the counter property of $B$. It is possible for an element $x$ to be such that $m^+(x) \neq 0$ and $m^-(x) \neq 0$ when membership function of the property overlaps that of its counter property over some portion of $X$. For the sake of simplicity, we shall use the symbol $B = (m^+, m^-)$ for the bipolar fuzzy set $B = \{(x, m^+(x), m^-(x)) \mid x \in X\}$.

For every two bipolar fuzzy sets $A = (m_A^+, m_A^-)$ and $B = (m_B^+, m_B^-)$ on $X$,

$(A \cap B)(x) = (min(m_A^+(x), m_B^+(x)), max(m_A^-(x), m_B^-(x)))$.

$(A \cup B)(x) = (max(m_A^+(x), m_B^+(x)), min(m_A^-(x), m_B^-(x)))$.

A *bipolar fuzzy graph* [1] with an underlying set $V$ is defined to be the pair $G = (A, B)$ where $A = (m_A^+, m_A^-)$ is a bipolar fuzzy set on $V$ and $B = (m_B^+, m_B^-)$ is a bipolar fuzzy set on $E \subseteq V \times V$ such that $m_B^+(x, y) \leq min\{m_A^+(x), m_A^+(y)\}$ and $m_B^-(x, y) \geq max\{m_A^-(x), m_A^-(y)\}$ for all $(x, y) \in E$. Here $A$ is called bipolar fuzzy vertex set of $V$, $B$ the bipolar fuzzy edge set of $E$. A bipolar fuzzy graph $G = (A, B)$ is said to be strong if $m_B^+(x, y) = min(m_A^+(x), m_A^+(y))$ and $m^-(x, y) = max(m_A^-(x), m_A^-(y))$. The *complement* [1] of a strong bipolar fuzzy graph $G$ is $\overline{G} = (\overline{A}, \overline{B})$ where $\overline{A} = (\overline{m}_A^+, \overline{m}_A^-)$ is a bipolar fuzzy set on $\overline{V}$ and $\overline{B} = (\overline{m}_B^+, \overline{m}_B^-)$ is a bipolar fuzzy set on $\overline{E} \subseteq \overline{V} \times \overline{V}$ such that

(1) $\overline{V} = V$,
(2) $\overline{m}_A^+(x) = m_A^+(x)$ and $\overline{m}_A^-(x) = m_A^-(x)$ for all $x \in V$,
(3)
$$\overline{m}_B^+(x, y) = \begin{cases} 0, & \text{if } m_B^+(x,y) > 0, \\ m_A^+(x) \wedge m_A^+(y), & \text{otherwise.} \end{cases}$$

$$\overline{m}_B^-(x, y) = \begin{cases} 0, & \text{if } m_B^-(x,y) < 0, \\ m_A^-(x) \vee m_A^-(y), & \text{otherwise.} \end{cases}$$

**Definition 1** *[2] Let $G = (A, B)$ be a bipolar fuzzy graph where $A = (m_1^+, m_1^-)$ and $B = (m_2^+, m_2^-)$ be two bipolar fuzzy sets on a non-empty finite set $V$ and $E \subseteq V \times V$ respectively. The graph $G$ is called complete bipolar fuzzy graph if $m_2^+(u,v) = min\{m_1^+(u), m_1^+(v)\}$ and $m_2^-(u,v) = max\{m_1^-(u), m_1^-(v)\}$ for all $u, v \in V$.*

**Definition 2** *[2] Let $G = (A, B)$ be a bipolar fuzzy graph where $A = (m_1^+, m_1^-)$ and $B = (m_2^+, m_2^-)$ be two bipolar fuzzy sets on a non-empty finite set $V$ and $E \subseteq V \times V$ respectively. If $d^+(u) = k_1, d^-(u) = k_2$ for all $u \in V$, $k_1, k_2$ are two real numbers, then the graph is called $(k_1, k_2)$-regular bipolar fuzzy graph.*

**Definition 3** *[2] Let $G = (A, B)$ be a bipolar fuzzy graph where $A = (m_1^+, m_1^-)$ and $B = (m_2^+, m_2^-)$ be two bipolar fuzzy sets on a non-empty finite set $V$ and $E \subseteq V \times V$ respectively. The total degree of a vertex $u \in V$ is denoted by $td(u)$ and defined as $td(u) = (td^+(u), td^-(u))$ where $td^+(u) = \sum_{(u,v) \in E} m_2^+(u,v) + m_1^+(u), td^-(u) = \sum_{(u,v) \in E} m_2^-(u,v) + m_1^-(u)$. If all the vertices of a bipolar fuzzy graph are of total degree, then the graph is said to be totally regular bipolar fuzzy graph.*

## 3 Some definitions related to bipolar fuzzy graphs

Degree of a vertex of a bipolar fuzzy graph is defined below.

**Definition 4** *Let $G = (A, B)$ be a bipolar fuzzy graph where $A = (m_1^+, m_1^-)$ and $B = (m_2^+, m_2^-)$ be two bipolar fuzzy sets on a non-empty finite set $V$ and $E \subseteq V \times V$ respectively. The positive degree of a vertex $u \in G$ is $d^+(u) = \sum_{(u,v) \in E} m_2^+(u,v)$. Similarly negative degree of a vertex $u \in G$ is $d^-(u) = \sum_{(u,v) \in E} m_2^-(u,v)$. The degree of a vertex $u$ is $d(u) = (d^+(u), d^-(u))$.*

**Example 1** We consider a bipolar fuzzy graph shown in Figure 0. Here $d^+(v_1) = 0.4 + 0.5 = 0.9, d^-(v_1) = (-0.3) + (-0.2) = -0.5$. So $d(v_1) = (0.9, -0.5)$. Similarly $d(v_2) = (0.9, -0.7)$ and $d(v_3) = (1, -0.8)$.

Order and size of a bipolar fuzzy graph is an important term in bipolar fuzzy graph theory. They are defined below.

**Definition 5** *Let $G = (A, B)$ be a bipolar fuzzy graph where $A = (m_1^+, m_1^-)$ and $B = (m_2^+, m_2^-)$ be two bipolar fuzzy sets on a non-empty finite set $V$ and $E \subseteq V \times V$ respectively. The order of $G$ is denoted by $O(G)$ and defined as $O(G) = (O^+(G), O^-(G))$ where $O^+(G) = \sum_{u \in V} m_1^+(u)$ and $O^-(G) = \sum_{u \in V} m_1^-(u)$.*

**Definition 6** Let $G = (A, B)$ be a bipolar fuzzy graph where $A = (m_1^+, m_1^-)$ and $B = (m_2^+, m_2^-)$ be two bipolar fuzzy sets on a non-empty finite set $V$ and $E \subseteq V \times V$ respectively. The size of $G$ is defined by $S(G) = (S^+(G), S^-(G))$ where $S^+(G) = \sum_{(u,v) \in E, u \neq v} m_2^+(u, v)$, $S^-(G) = \sum_{(u,v) \in E, u \neq v} m_2^-(u, v)$.

**Example 2** In Figure 1, a bipolar fuzzy graph is shown. Here $O(G) = (1.9, -1.4)$ and $S(G) = (1.4, -1)$.

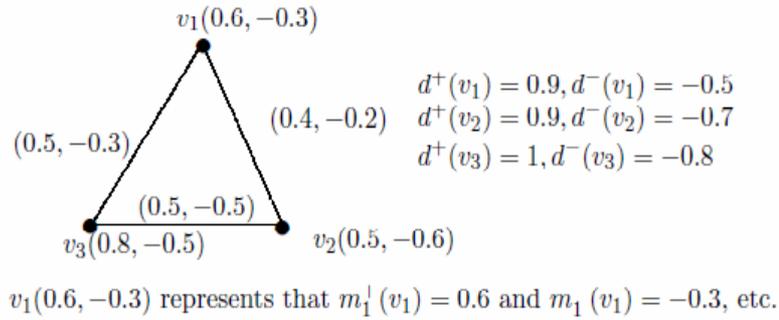

Figure 1: Degrees of the vertices of a bipolar fuzzy graph.

**Definition 7** Let $G = (A, B)$ be a bipolar fuzzy graph where $A = (m_1^+, m_1^-)$ and $B = (m_2^+, m_2^-)$ be two bipolar fuzzy sets on a non-empty finite set $V$ and $E \subseteq V \times V$ respectively. The underlying crisp graph of $G$ is the crisp graph $G' = (V', E')$ where $V' = \{v \mid m_1^+(v) > 0 \text{ or } m_1^-(v) < 0\}$ and $E' = \{(u, v) \mid m_2^+(u, v) > 0 \text{ or } m_2^-(u, v) < 0\}$.

**Definition 8** A bipolar fuzzy graph is said to be connected if its underlying crisp graph is connected.

**Example 3** A crisp underlying graph is shown in Figure 2. It is also an example of connected bipolar fuzzy graph.

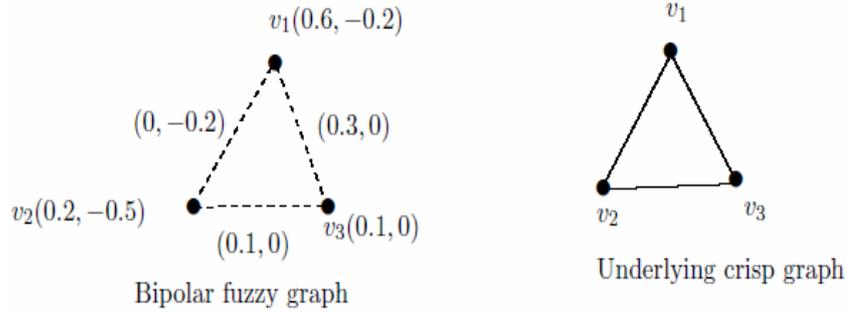

**Figure 2: Example of connected regular bipolar fuzzy graph.**

**Theorem 1** *Let $G$ be a regular bipolar fuzzy graph where induced crisp graph $G'$ is an even cycle. Then $G$ is regular bipolar fuzzy graph if and only if either $m_2^+$ and $m_2^-$ are constant functions or alternate edges have same positive membership values and negative membership values.*

**Proof.** Let $G = (A, B)$ be a regular bipolar fuzzy graph where $A = (m_1^+, m_1^-)$ and $B = (m_2^+, m_2^-)$ be two bipolar fuzzy sets on a non-empty finite set $V$ and $E \subseteq V \times V$ respectively and underlying crisp graph $G'$ of $G$ be an even cycle. If either $m_2^+, m_2^-$ are constant functions or alternate edges have same positive and negative membership values, then $G$ is a regular bipolar fuzzy graph. Conversely, suppose $G$ is a $(k_1, k_2)$-regular bipolar fuzzy graph. Let $e_1, e_2, \ldots, e_n$ be the edges of $G'$ in order. As in the theorem 3,

$$m_2^+(e_i) = \begin{cases} c_1, & \text{if } i \text{ is odd}, \\ k_1 - c_1, & \text{if } i \text{ is even}. \end{cases}$$

$$m_2^-(e_i) = \begin{cases} c_2, & \text{if } i \text{ is odd}, \\ k_2 - c_2, & \text{if } i \text{ is even}. \end{cases}$$

If $c_1 = k_1 - c_1$, then $m_2^+$ is constant. If $c_1 \neq k_1 - c_1$, then alternate edges have same positive and negative membership values. Similarly for $m_2^-$. Hence the results. □

**Theorem 2** *The size of a $(k_1, k_2)$-regular bipolar fuzzy graph is $(\frac{pk_1}{2}, \frac{pk_2}{2})$ where $p = |V|$.*

**Proof.** Let $G = (A, B)$ be a bipolar fuzzy graph where $A = (m_1^+, m_1^-)$ and $B = (m_2^+, m_2^-)$ be two bipolar fuzzy sets on a non-empty finite set $V$ and $E \subseteq V \times V$ respectively. The size of $G$ is $S(G) = (\sum_{u \neq v} m_2^+(u, v), \sum_{u \neq v} m_2^-(u, v))$. We have $\sum_{v \in V} d(v) = 2[\sum_{(u,v) \in E} m_2^+(u, v), \sum_{(u,v) \in E} m_2^-(u, v)] = 2S(G)$. So $2S(G) = \sum_{v \in V} d(v)$. i.e $2S(G) = (\sum_{v \in V} k_1, \sum_{v \in V} k_2)$. This gives $2S(G) = (pk_1, pk_2)$. Hence the result. □

**Theorem 3** *If $G$ is $(k,k')$-totally regular bipolar fuzzy graph, then $2S(G)+O(G)=(pk,pk')$ where $p=|V|$.*

**Proof.** Let $G=(A,B)$ be a bipolar fuzzy graph where $A=(m_1^+,m_1^-)$ and $B=(m_2^+,m_2^-)$ be two bipolar fuzzy sets on a non-empty finite set $V$ and $V\times V$ respectively. Since $G$ is a $(k,k')$-totally regular fuzzy graph. So $k=td^+(v)=d^+(v)+m_1^+(v)$ and $k'=td^-(v)=d^-(v)+m_1^-(v)$ for all $v\in V$. Therefore $\sum_{v\in V}k=\sum_{v\in V}d^+(v)+\sum_{v\in V}m_1^+(v)$ and $\sum_{v\in V}k'=\sum_{v\in V}d^-(v)+\sum_{v\in V}m_1^-(v)$. $pk=2S^+(G)$ and $pk'=2S^-(G)$. So $pk+pk'=2(S^+(G)+S^-(G))+O^+(G)+O^-(G)$. Hence $2S(G)+O(G)=(pk,pk')$. □

## 4 Irregular bipolar fuzzy graphs

Irregular bipolar fuzzy graphs are as important as regular bipolar fuzzy graphs. We now define it.

**Definition 9** *Let $G=(A,B)$ be a bipolar fuzzy graph where $A=(m_1^+,m_1^-)$ and $B=(m_2^+,m_2^-)$ be two bipolar fuzzy sets on a non-empty finite set $V$ and $E\subseteq V\times V$ respectively. $G$ is said to be irregular bipolar fuzzy graph if there exists a vertex which is adjacent to a vertex with distinct degrees.*

**Example 4** Let $G=(A,B)$ be a bipolar fuzzy graph where $A=(m_1^+,m_1^-)$ and $B=(m_2^+,m_2^-)$ be two bipolar fuzzy sets on a non-empty finite set $V$ and $E\subseteq V\times V$ respectively, where $V=\{v_1,v_2,v_3,v_4\}$. $d(v_1)=(0.8,-1)$, $d(v_2)=(0.8,-1)$, $d(v_3)=(1.2,-1.4)$, $d(v_4)=(0.4,-0.4)$. Here $d(v_2)\neq d(v_3)$. So this graph is an example of irregular bipolar fuzzy graph, shown in Figure 3.

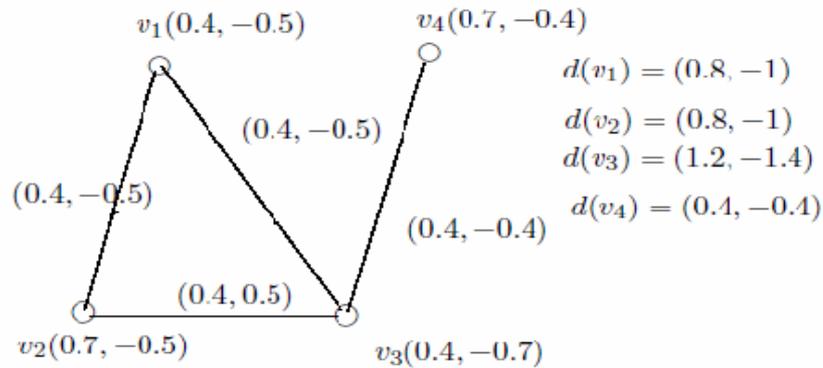

**Figure 3: Example of irregular bipolar fuzzy graph.**

Neighbourly irregular bipolar fuzzy graph is a special case of irregular bipolar fuzzy graph.

**Definition 10** *Let $G$ be a connected bipolar fuzzy graph. Then $G$ is called neighbourly irregular bipolar fuzzy graph if for every two adjacent vertices of $G$ have distinct degrees.*

**Definition 11** *Let $G = (A, B)$ be a bipolar fuzzy graph where $A = (m_1^+, m_1^-)$ and $B = (m_2^+, m_2^-)$ be two bipolar fuzzy sets on a non-empty finite set $V$ and $E \subseteq V \times V$ respectively. $G$ is said to be totally irregular bipolar fuzzy graph if there exists a vertex which is adjacent to a vertex with distinct total degrees.*

**Example 5** Let $G = (A, B)$ be a bipolar fuzzy graph where $A = (m_1^+, m_1^-)$ and $B = (m_2^+, m_2^-)$ be two bipolar fuzzy sets on a non-empty finite set $V$ and $E \subseteq V \times V$ respectively. Let $V = \{v_1(0.5, -0.4), v_2(0.4, -0.6), v_3(0.3, 0.2)\}$. $d(v_1) = (0.7, -1.1)$, $d(v_2) = (0.6, -1.3)$ and $d(v_3) = (0.5, -1.2)$. Here $td(v_1) = (1.2, -1.5)$ and $td(v_2) = (1, -1.9)$. So this graph is an example of totally irregular bipolar fuzzy graph.

**Definition 12** *Let $G$ be a connected bipolar fuzzy graph. Then $G$ is called neighbourly totally irregular bipolar fuzzy graph if for every two adjacent vertices of $G$ have distinct total degrees.*

**Definition 13** *Let $G$ be a connected bipolar fuzzy graph. Then $G$ is called highly irregular bipolar fuzzy graph if every vertex of $G$ is adjacent to vertices with distinct degrees.*

**Example 6** Let $G = (A, B)$ be a bipolar fuzzy graph where $A = (m_1^+, m_1^-)$ and $B = (m_2^+, m_2^-)$ be two bipolar fuzzy sets on a non-empty finite set $V$ and $E \subseteq V \times V$ respectively where $V = \{v_1, v_2, v_3\}$. $d(v_1) = (0.6, -0.5)$, $d(v_2) = (0.8, -0.4)$ and $d(v_3) = (0.5, -0.4)$. The graph is an example of highly irregular bipolar fuzzy graph.

A highly irregular bipolar fuzzy graph need not be neighbourly irregular bipolar fuzzy graph. As for example we consider a bipolar fuzzy graph of vertices
$v_1(0.5, -0.4), v_2(0.6, -0.5), v_3(0.5, -0.4), v_4(0.4, -0.4)$ with
$m_2^+(v_1, v_2) = 0.4, m_2^-(v_1, v_2) = -0.3, m_2^+(v_2, v_3) = 0.2, m_2^-(v_2, v_3) = -0.4$,
$m_2^+(v_2, v_4) = 0.2, m_2^-(v_2, v_4) = -0.4, m_2^+(v_3, v_4) = 0.4, m_2^-(v_3, v_4) = -0.3$. So $d(v_1) = (0.4, -0.3), d(v_2) = (0.8, -1.1), d(v_3) = (0.6, -0.7), d(v_4) = (0.6, -0.7)$. Here the bipolar fuzzy graph is highly irregular but not neighbourly irregular as $d(v_3) = d(v_4)$.

**Theorem 4** *Let $G$ be a bipolar fuzzy graph. Then $G$ is highly irregular bipolar fuzzy graph and neighbourly irregular bipolar fuzzy graph if and only if the degrees of all vertices of $G$ are distinct.*

**Proof.** Let $G = (A, B)$ be a bipolar fuzzy graph where $A = (m_1^+, m_1^-)$ and $B = (m_2^+, m_2^-)$ be two bipolar fuzzy sets on a non-empty finite set $V$ and $V \times V$ respectively. Let $V = \{v_1, v_2, \ldots, v_n\}$. We assume that $G$ is highly irregular and neighbourly irregular bipolar fuzzy graphs. Let the adjacent vertices of $u_1$ be $u_2, u_3, \ldots, u_n$ with degrees $(k_2^+, k_2^-), (k_3^+, k_3^-), \ldots (k_n^+, k_n^-)$ respectively. As $G$ is highly and neighbourly irregular, $d(u_1) \neq d(u_2) \neq d(u_3) \neq \ldots \neq d(u_n)$. So it is obvious that all vertices are of distinct degrees.

Conversely, assume that the degrees of all vertices of $G$ are distinct. This means that every two adjacent vertices have distinct degrees and to every vertex the adjacent vertices have distinct degrees. Hence $G$ is neighbourly irregular and highly irregular fuzzy graphs. □

Complement of a bipolar fuzzy graph is defined below.

**Definition 14** *Let $G = (A, B)$ be a bipolar fuzzy graph where $A = (m_1^+, m_1^-)$ and $B = (m_2^+, m_2^-)$ be two bipolar fuzzy sets on a non-empty finite set $V$ and $E \subseteq V \times V$ respectively. The complement of a bipolar fuzzy graph $G$ is $\overline{G} = (\overline{A}, \overline{B})$ where $\overline{A} = (\overline{m}_A^+, \overline{m}_A^-)$ is a bipolar fuzzy set on $\overline{V}$ and $\overline{B} = (\overline{m}_B^+, \overline{m}_B^-)$ is a bipolar fuzzy set on $\overline{E} \subseteq \overline{V} \times \overline{V}$ such that*

(1) $\overline{V} = V$,

(2) $\overline{m}_A^+(x) = m_A^+(x)$ and $\overline{m}_A^-(x) = m_A^-(x)$ for all $x \in V$,

(3)
$$\overline{m}_B^+(x, y) = \begin{cases} 0, & \text{if } m_B^+(x, y) > 0, \\ m_A^+(x) \wedge m_A^+(y), & \text{otherwise.} \end{cases}$$

$$\overline{m}_B^-(x, y) = \begin{cases} 0, & \text{if } m_B^-(x, y) < 0, \\ m_A^-(x) \vee m_A^-(y), & \text{otherwise.} \end{cases}$$

If a bipolar fuzzy graph $G$ is neighbourly irregular, then $\overline{G}$ is not be neighbourly irregular. Let us consider a bipolar fuzzy graph whose non adjacent vertices are of same degree. Then clearly, adjacent vertices of $\overline{G}$ are of same degree. Hence the statement is true.

**Theorem 5** *Let $G$ be a bipolar fuzzy graph. If $G$ is neighbourly irregular and $m_1^+, m_1^-$ are constant functions, then $G$ is a neighbourly total irregular bipolar fuzzy graph.*

**Proof.** Let $G = (A, B)$ be a bipolar fuzzy graph where $A = (m_1^+, m_1^-)$ and $B = (m_2^+, m_2^-)$ be two bipolar fuzzy sets on a non-empty finite set $V$ and $E \subseteq V \times V$ respectively. Assume that $G$ is a neighbourly irregular bipolar fuzzy graph. i.e. the degrees of every two adjacent vertices are distinct. Consider two adjacent vertices $u_1$ and $u_2$ with distinct degrees $(k_1^+, k_1^-)$ and $(k_2^+, k_2^-)$ respectively. Also let, $m_1^+(u) = c_1$ for all $u \in V$, $m_1^-(u) = c_2$ for all $u \in V$ where $c_1 \in (0, 1], c_2 \in [-1, 0)$ are constant. Therefore
$td(u_1) = (d^+(u_1) + c_1, d^-(u_1) + c_2) = (k_1^+ + c_1, k_1^- + c_2)$,
$td(u_2) = (d^+(u_2) + c_1, d^-(u_2) + c_2) = (k_2^+ + c_1, k_2^- + c_2)$. Clearly $td(u_1) \neq td(u_2)$. Therefore for any

two adjacent vertices $u_1$ and $u_2$ with distinct degrees, its total degrees are also distinct, provided $m_1^+, m_1^-$ are constant functions. The above argument is true for every pair of adjacent vertices in $G$. □

**Theorem 6** *Let $G$ be a bipolar fuzzy graph. If $G$ is neighbourly total irregular and $m_1^+, m_1^-$ are constant functions, then $G$ is a neighbourly irregular bipolar fuzzy graph.*

**Proof.** Let $G = (A, B)$ be a bipolar fuzzy graph where $A = (m_1^+, m_1^-)$ and $B = (m_2^+, m_2^-)$ be two bipolar fuzzy sets on a non-empty finite set $V$ and $V \times V$ respectively. Assume that $G$ is a neighbourly total irregular bipolar fuzzy graph. i.e. the total degree of every two adjacent vertices are distinct. Consider two adjacent vertices $u_1$ and $u_2$ with degrees $(k_1^+, k_1^-)$ and $(k_2^+, k_2^-)$ respectively. Also assume that $m_1^+(u) = c_1$ for all $u \in V$, $m_1^-(u) = c_2$ for all $u \in V$ where $c_1 \in (0,1], c_2 \in [-1,0)$ are constants. Also $td(u_1) \neq td(u_2)$. We are to prove that $d(u_1) \neq d(u_2)$. $k_1^+ + c_1 \neq k_2^+ + c_1$ and $k_1^- + c_2 \neq k_2^- + c_2$. So $k_1^+ \neq k_2^+$ and $k_1^- \neq k_2^-$. Hence the degrees of adjacent vertices of $G$ are distinct. This is true for every pair of adjacent vertices in $G$. Hence the result. □

## 5 Conclusion

In this paper we have described degree of a vertex, order, size and underlying crisp graph of a bipolar fuzzy graphs. The necessary and sufficient conditions for a bipolar fuzzy graph to be the regular bipolar fuzzy graphs have been presented. Size of a bipolar fuzzy graphs and relation between size and order of a bipolar fuzzy graphs have been calculated. We have defined irregular bipolar fuzzy graphs, neighbourly irregular, totally and highly irregular bipolar fuzzy graphs. Some relations about the defined graphs have been proved.

Some theoretical discussions on bipolar fuzzy graphs have been made. We want to make, in near future, some algorithm and models using these results.

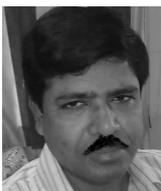
**Madhumangal Pal** is a Professor of Applied Mathematics with Oceanology and Computer Programming, Vidyasagar University, India. He is Editor-in-Chief of *Journal of Physical Sciences* and member of Editorial Board of *International Journal of Computer Sciences,* of *Systems Engineering and Information Technology,* of *International Journal of Fuzzy Systems & Rough Systems,* of *Advanced Modeling and Optimization, Romania* and of *International Journal of Logic and Computation, Malaysia.* He is *a* reviewer of several international journals. He has written several books on Mathematics and Computer Science. His research interest includes computational graph theory, fuzzy matrices, game theory and regression analysis, parallel and genetic algorithms, etc.

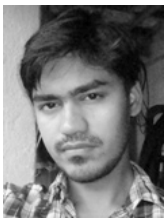
**Sovan Samanta** received his B.Sc. degree in 2007 and M.Sc. degree in 2009 in Applied Mathematics from Vidyasagar University, India. He is now currently a research scholar in the Department of Applied Mathematics, Vidyasagar University since 2010. His research interest includes fuzzy graph theory and bipolar fuzzy graph theory .